%% file: elsarticle-template-num-names.tex
\documentclass[preprint,12pt]{elsarticle}
\usepackage{graphicx}
\usepackage{algorithmic}
\usepackage[ruled,vlined]{algorithm2e}

\usepackage{amssymb}

\usepackage{lineno}
\journal{Blockchain: Research and Applications}

\begin{document}

\begin{frontmatter}



\title{A Collaboration Strategy in the Mining Pool for Proof-of-Neural-Architecture Consensus}


\author[inst1]{Boyang Li}

\affiliation[inst1]{organization={University of Notre Dame},
            addressline={Department of Computer Science and Engineering}, 
            city={Notre Dame},
            postcode={46556}, 
            state={Indiana},
            country={USA}}

\author[inst1]{Qing Lu}
\author[inst2]{Weiwen Jiang}
\author[inst1]{Taeho Jung}

\author[inst1]{Yiyu Shi}

\affiliation[inst2]{organization={George Mason University},
            addressline={Department of Electrical and Computer Engineering}, 
            city={Fairfax},
            postcode={22030}, 
            state={Virginia},
            country={USA}}
            
\begin{abstract}

In most popular public accessible cryptocurrency systems, the mining pool plays a key role because mining cryptocurrency with the mining pool turns the non-profitable situation into profitable for individual miners. In many recent novel blockchain consensuses, the deep learning training procedure becomes the task for miners to prove their workload, thus the computation power of miners will not purely be spent on the hash puzzle. In this way, the hardware and energy will support the blockchain service and deep learning training simultaneously. While the incentive of miners is to earn tokens, individual miners are motivated to join mining pools to become more competitive. In this paper, we are the first to demonstrate a mining pool solution for novel consensuses based on deep learning. 

The mining pool manager partitions the full searching space into subspaces and all miners are scheduled to collaborate on the Neural Architecture Search (NAS) tasks in the assigned subspace. Experiments demonstrate that the performance of this type of mining pool is more competitive than an individual miner. Due to the uncertainty of miners' behaviors, the mining pool manager checks the standard deviation of the performance of high reward miners and prepares backup miners to ensure completion of the tasks of high reward miners. 
\end{abstract}



\begin{keyword}
Blockchain \sep Consensus \sep Deep learning \sep Mining Pool \sep Neural architecture search(NAS)
\PACS 0000 \sep 1111
\MSC 0000 \sep 1111
\end{keyword}

\end{frontmatter}


\input{body}




\bibliographystyle{elsarticle-num-names} 
\bibliography{bib}





\end{document}

%% file: body.tex
\section{Introduction}\label{sec:introduction}

Previously, multiple publications described novel blockchain consensuses that support alternative mining puzzles other than the hash algorithm. As a result, the computation power of miners' hardware will not be wasted on a pure brute-force algorithm. Especially, Privacy‐Preserving Blockchain Mining \cite{turesson2018deep}, Coin.AI~\cite{Gmez2019CoinAIAP}, WekaCoin~\cite{bravo2019proof}, DLBC \cite{Li_2019_CVPR_Workshops, li2019dlbc}, and PoDL \cite{chenli2019energy} are on top of novel consensus which perform deep learning training algorithms as proof-of-useful-work (PoUW).

Deep learning (DL) has developed rapidly in recent decades and has been widely applied in different fields. Training a deep learning model with good performance not only takes a massive amount of energy but also requires significant research effort. Neural architecture search (NAS) has recently become popular because it can help researchers to design DL models automatically. Similar to blockchain services, NAS also requires enormously high computation power, and insufficient computation resources will slow down the productivity of researchers. With the help of novel consensuses, the computation power of miners in blockchain services could be leveraged to accelerate NAS.

In the permission-less blockchain system, the security of the system depends on the high volume of individual miners. The incentive to mine is to earn tokens. For an individual miner, it is profitable only if the value of the earned token is more than the electricity bill. Because only the winner miner will receive a block reward, it is a risk for individual miners that they have spent electricity but do not receive the block reward. In practice, the majority of the miners will not receive the block reward. The possibility for an individual miner to earn the credit is the ratio of its computation power to the computation power of the whole network in PoW consensus. In a mature permission-less blockchain system, the computation power of the entire network is ideally exceptionally high, so it can serve as the backup of the security property of the blockchain system. Therefore, the possibility for an individual miner to earn the reward is extremely low. 

For nowadays publicly accessible cryptocurrencies, an individual miner most likely will join a mining pool to earn tokens. Because mining with the mining pool will minimize the risk of zero earnings in a short term. A mining pool is formed by a group of miners and it behaves as mutual insurance\cite{narayanan2016bitcoin}. But, the mining pool for hash-based consensus is different from the PoNAS consensus. In this paper we proposed a collaboration strategy in the mining pool for PoNAS consensus. To my best knowledge, we are the first to demonstrate the design of the mining pool for NAS-based PoUW.

The main contribution of this design is as followed:
\begin{itemize}
\item 
We explained the benefits of the mining pool for the PoNAS and discussed the difficulties to achieve the mining pool. An intuitive method is to adopt the existing distributed deep learning frameworks to train the PoDL workload. Without optimization of scheduling of the training tasks, the slowest miner will become the bottleneck of overall performance and most of the miners will be idle. 
\item 
We introduced a collaboration strategy of exploration and exploitation to the mining pool. Therefore, the miners with less computation power will not drawback the overall performance, and the resource of these miners will not be wasted.  

\item 
We applied the naive parallel computing solution with which the dependency between miners is further reduced. The bottleneck of the network does not exist in the design. 
\end{itemize}

In addition, this design leveraged the NAS as the workload. The actual training tasks among miners are different while the final proof information is the same, which is an effective neural architecture. This is different from  PoUW and PoDL, which require all miners to work on the same workload within one block. The effectiveness of NAS is beyond the contribution of this paper, yet we gave an example that miners can train the DL model with different architectures while providing the information to be proven by consensus. 


\section{Background \& Related Work} \label{sec:background}
\subsection{Consensus of existing cryptocurrency}

The Bitcoin is built on top of the Proof of Work (PoW) consensus which requests all miners to solve problems. Generally, the required problems in PoW consensus are easy to validate and hard to solve. The PoW is stable but its power consumption is enormous.  
The Ethereum is a popular cryptocurrency based on Proof of Stake (PoS) consensus which decides the creator of the new block. In PoS, the amount of owned token of a certain miner will prove the support the authority. In this consensus, the computation is relatively more efficient than PoW, but it is less stable and robust owing to various limitations.\cite{poelstra2014distributed,ogawaproposal}. 

\subsection{Proof-of-Deep-Learning (PoDL):}
In the consensus based on the deep learning algorithm, it divides each block time into two or more interval phases \cite{chenli2019energy, Li_2019_CVPR_Workshops, li2019dlbc}. In general, it includes the initialization phase, the training phase, and the validation phase. 

In the initialization phase,  all miners confirm the target task and evaluate the training setup, such as the target training epochs and the size of the dataset. In the training phase, miners train the confirmed target task and commit their model before the training phase ends. Here the miners submit the hash of their deep learning model, training results, and miners ID. The task publishers release training dataset and deep learning training source code. \cite{chenli2019energy, Li_2019_CVPR_Workshops, li2019dlbc}

In the validation phase, the task publisher releases the test dataset to miners and full nodes, and each miner submits (1) the block header and the block that contains information describing the trained model on top of existing attributes, (2) the trained model, and (3) the accuracy of the trained model, to full nodes. The full nodes validate the submitted models. Here, full nodes discard all submissions if the model was not committed before the training phase ends in the current block (\textit{i.e.,} hash of the model and ID have not been received), thus it prevents from miners over-fitting their models on the disclosed test dataset or stealing others models. 
The full nodes will confirm the new block and authorize the creator of this block. During the confirmation process, the full nodes validate the model with the highest accuracy amount of all submissions. If the validated results equal as claimed, the confirmation is finished. Otherwise, the full nodes will continue the process and validate the models in decreasing order of the claimed accuracy. This confirmation process yields a robust consensus. \cite{chenli2019energy, Li_2019_CVPR_Workshops, li2019dlbc}

\subsection{Other Proof-of-Useful-Work mechanisms:}
Primecoin \cite{king2013primecoin} is built on top of Proof-of-Useful-Work mechanisms that request all miners to solve problems. Here, the puzzle is to find a special sequence of prime numbers. The consensus helps to solve mathematical problems, \textit{i.e.,} discovering the Cunningham chain. 
Hybrid mining ~\cite{chatterjee2019hybrid} solves problems with the computational power of the blockchain system. 
Privacy‐Preserving Blockchain\cite{turesson2018deep} introduced their two
parallel chains and dynamic committee members' strategy. Here, PoUW runs on top of the long-interval and transactions in short-interval. 
In Coin.AI~\cite{Gmez2019CoinAIAP}, miners will train DL model mode and also store the valuable data for tokens. 
WekaCoin~\cite{bravo2019proof} will contribute to creating a public distributed and verified database. 

\begin{figure*}[h]
\begin{center}
    \includegraphics[width=1\columnwidth]{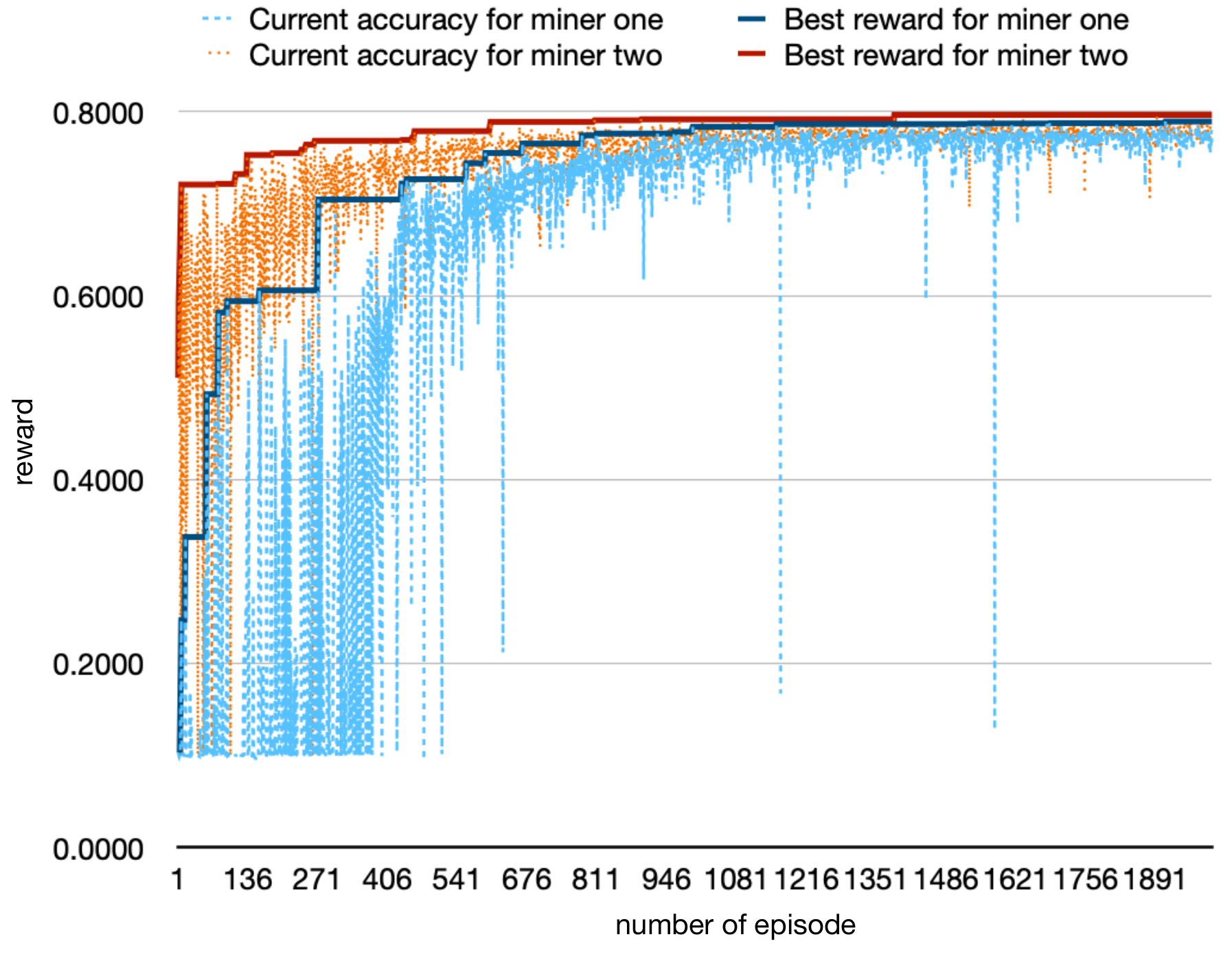}
\end{center} 
    \caption{The full results of NAS of two miners searching in different searching spaces. The solid lines shows the best reward of NAS in two spaces. The dash lines shows the current reward of each configuration during searching procedure. The x-axis is the number of episodes and y-axis is the reward.}
\label{fig:ts}
\end{figure*}

\subsection{NAS}
While artificial neural networks have achieved ground-breaking success in machine learning applications, their topology is trending towards complex and indecipherable. As a result, the traditional architecture engineering process relying on the labor and expertise of human designers are regarded as either impossible or inefficient in pushing the state of the art to the next level. Therefore, design automation in neural networks is logically the future of machine learning. In the Fig. \ref{fig:ts}, it shows the full results of the NAS of two miners searching in different searching spaces. The solid lines show the best reward of NAS in two spaces. The dash lines show the current reward of each configuration during the searching procedure.

As an essential part of AutoML, neural architecture search refers to the technique of automating exploration in the design space of neural network topology. A typical use case of NAS would be having an abstract structure with multiple sub-structures to be optimized using a set of optional building blocks. 
This design space is too huge to be exhausted and manually pruned, so the success of NAS is dependent on a carefully devised searching strategy. In order to find improved architectures over the previous models, a variety of search algorithms have been studied in existing works, including reinforcement learning \cite{zoph2016neural_replace, zoph2017learning_replace}, evolutionary methods \cite{real2017largescale_replace, liu2017hierarchical_replace}, Bayesian Optimization \cite{k2018neural_replace}, hill-climbing \cite{elsken2017simple_replace}, etc. It is noted that NAS is a computational-intensive problem, so how to formulate the NAS problem to improve efficiency has attracted more and more research interest \cite{pham2018efficient_replace, yan2019hmnas_replace}.

A mainstream NAS framework of today is the weight sharing scheme where multiple child networks are trained together as one ``supernet'' and inherit the same set of weights for evaluation \cite{guo2019single_replace}. In such frameworks, the ``Childnet's'' are sampled by some searching algorithm for optimal performance. Particularly, \cite{liu2018darts_replace} proposed the gradient-based method named ``DARTS'' to jointly search and train the child networks with very high efficiency. Soon after NAS was brought up, it was applied under hardware constraints to form a research topic\textemdash hardware-aware NAS \cite{FNAS_replace, srinivas2019hardware_replace}. Since quantization is one of the most common techniques for approximated computation in hardware, some works employed NAS to search the best configuration for quantizing neural networks, a.k.a quantization-aware NAS \cite{wang2018haq_replace}. Hardware-aware NAS is not only deployment-oriented but also efficient as the hardware search can be guided.

\section{Methodology} \label{sec:methodology}
\subsection{Overview of mining pool} \label{sec:sub:overview_mining_pool}
In this mining pool, the participants include miners and managers. 
The pool manager will receive the rewards once any of the participants find the block and the manager will distribute the rewards to each participant \cite{narayanan2016bitcoin}.  
The pool manager normally hosts very powerful servers to maintain a stable connection and job distribution. 
Miners pay mining fees as a return to the good performing pool manager. 
The miners in the pool will contribute to the assigned tasks.

For the consensus based on hash, the miners are working independently thus the individual performance of a miner will not suffer from dependency from miners. 
The amount of the reward distributions will depend on the ratio of computation power and contribution.
The job distribution is relatively simple because the computation is relatively independent.

In the recent deep learning-based PoUW consensus \cite{chenli2019energy, Li_2019_CVPR_Workshops, li2019dlbc, turesson2018deep, Gmez2019CoinAIAP, bravo2019proof, qu2021proof}, multiple papers discussed the possible solution that the computation power of miners will be spent on relative useful work and here are the deep learning training tasks. For instance, PoDL \cite{chenli2019energy}, Proof of Federated Learning \cite{qu2021proof}, and etc.. 
Because all miners will wish to avoid risks, a mining pool will appear naturally.

In the mining pool based on NAS, the amount of the reward distributions will still depend on the ratio of computation power and contribution. 
But a weak miner may hold back the performance of the whole mining pool. 
Therefore, the Intuitive solution is that the manager will distribute easier jobs to weak miners and distribute harder jobs to strong miners. Therefore, all miners will be able to deliver useful results
We will introduce miner type in the Sec. \ref{sec:sub:sub:miner_types}

\subsection{Design of mining pool}\label{sec:sub:Design of mining pool}
\subsubsection{Deep learning consensus}
As introduced in Sec.\ref{sec:background}, there are multiple existing publications about a novel blockchain consensus based on a deep learning algorithm instead of a brute force algorithm. For demonstration purposes, this mining pool will service a consensus such as Deep Learning consensus \cite{Li_2019_CVPR_Workshops, li2019dlbc}. 

In this consensus \cite{Li_2019_CVPR_Workshops, li2019dlbc}, it described the design of three phases block interval and task scheduling for each interval phase. For phase one, it is the initial stage that miners will confirm the target task to train and evaluate the difficulty of the task. In \cite{Li_2019_CVPR_Workshops, li2019dlbc}, the difficulties of the tasks include model size, data size, network bandwidth, FLOPs of the task, and computation power of the network. For phase two, it is the time for GPUs of miners to train the DL tasks and full nodes to spread submitted tasks. Miner nodes select the next target task to train based on the ranking score. The ranking score is the ratio of the difficulty of the task over the task reward. All miner nodes will only allow submitting the training results before phase two is finished. For phase three, full nodes rank the submitted training results and evaluated the winner model. During phase two and phase three, once a task is selected for the next block, all miners will fetch the data from full nodes and task publishers. Once the performance of the winner model is the same as the miner claimed, the winner miner will generate the current block and full nodes start to spread the current block. If the performance of the winner model is worse than the submitted value, the full nodes will remove it and evaluate the next best model. For all nodes, if they find the block is generated, they will validate and confirm the block. Once a block is confirmed for an individual node, it will move onto phase one of the next block. 

For a mining pool, the pool manager will split the searching space of the selected NAS task into multiple subspaces. Therefore, the whole mining pool may achieve better performance. 


\subsubsection{Miner types} \label{sec:sub:sub:miner_types}
In the mining pool design, we separate miners into strong miners and weak miners. For strong miners, they can finish the search task in a given subspace. For weak miners, they cannot finish the search task to give subspace due to the limitation of network bandwidths, hardware, etc.. 

In practice, the performance of a weak miner would affect the performance of the whole mining pool. In the experiments, we noticed that some subspace can be better than others that the miner will find a better neural architecture in this ``lucky space''. In the case, if this ``lucky space'' is assigned to a weak miner and the miner may waste the good opportunity to find the better neural architecture in this subspace. To increase the efficiency of the mining pool, the pool manager will try to reduce the overlap between each subspace. Therefore, it is very possible that a mining pool will not find this neural architecture and the performance of the pool will be held back.



\subsubsection{General embarrassingly parallel computations} \label{sec:sub:sub:parallelism}

\begin{algorithm*}[ht]
\SetAlgoLined
\textbf{Input:}
\\Hyperparameters $h_1, h_2, ..., h_n$ with $h_i$ of searching range $r_i = \{R_i^1, R_i^2, ..., R_i^k\}$, $m$ is the total number of miners. 

\textbf{Return:}
\\A list of subspaces, Subspace = $[S_1, S_2, ..., S_m]$.
\SetAlgoLined

\textbf{Initialization:}
\\Initialize empty list with size equals to m\;
Initialize table T with the key equals to index of hyperparameters and the value equals to a list of searching range for the corresponding hyperparameter\;
\For{$i = 1;\ i \leq n;\ i = i + 1$}{
    T[$i$] equals to a list of all subsets of $r_i$\;
  }
  
\textbf{Partition:}
\\  \For{$i = 1;\ i \leq m;\ i = i + 1$}{
        Initialize the space $S_i$ for miner $i$\;
        \For{$j = 1;\ j \leq n;\ j = j + 1$}{
        The selected searching range for hyperparameter j = T[$j$][randint()])
        $S_i$.append(The selected searching range)
        }
    Subspace.append($S_i$)
  }
return Subspace\;
 \caption{Partition of search space for miners}
 \label{algo:subspace}
\end{algorithm*}

In this mining pool, the pool manager will split the search space into multiple subspaces and each miner will search the neural architecture independently. Therefore, it will reduce the effect of network bandwidth on the performance of whole searching tasks. This parallelism will not request any dependence during the search for each miner thus it is named embarrassingly parallelism. 


In NAS tasks, the searching space is a high dimension which covers all combination of neural network configuration. To search a high-performance model in this high dimension space may request many experiments. The NAS will help to find a good configuration of a neural network. 

In the algorithm \ref{algo:subspace}, each hyperparameter $h_i$ $(\in \{h_1, h_2, ..., h_n\})$ of the target neural network can be selected from a searching range $r_i = [R_i^1, R_i^2, ..., R_i^k]$. The $i$ is the index of the hyperparameter and $k$ represents the maximum size of the searching range of the $i$ hyperparameter, and $n$ is the total number of hyperparameter.
A subspace $S$ $(\in S_1, S_2, ..., S_m)$ formed by $n$ elements and each element formed by a searching range of one hyperparameter. For instance, to assign a subspace to a miner, the mining pool manager follows algorithm \ref{algo:subspace} and selects one searching range for each hyperparameter. In section \ref{sec:experiment}, we will demonstrate the algorithm in a certain value based on a given NAS task.

\subsubsection{Collaboration strategy of exploration and exploitation} \label{sec:sub:sub:exploration and exploiting}

The tasks for the pool manager include collecting tasks, splitting tasks, collecting results, and submitting the best results. The pool manager collects input information from full nodes and splits the searching space into subspaces as described in algorithm \ref{algo:subspace}. The pool manager collects the best results from all miners and submits the best solution to full nodes before the training phase ends. 

Here, the collaboration strategy of exploration and exploitation is to schedule strong miners and weak miners separately. In practice, some subspaces may be easier to find better performance. The NAS agent is designed to propose a certain neural architecture without training it. If a weak miner is assigned to search neural architecture in the subspace which contains the final best solution, the weak miner may not have sufficient computation power to find the final target architecture, thus it will waste the opportunity. 
Therefore, the searching space will be only sent to the strong miners. Once a solution is confirmed to beats the current best results, the strong miners will share the corresponding hyperparameter with weak miners. The weak miners will continue to exploit the confirmed architecture. The weak miners will also update the improved solution with strong miners. All miners submit the best solution to the pool manager once they find a solution that beats the current best results.

\section{Experiment} \label{sec:experiment}

\begin{table*}[ht]
\centering
\caption{Model architecture subspaces for subspaces S1 to S9 and full space.}
\resizebox{\columnwidth}{!}{%
    \begin{tabular}{|l|l|l|l|l|l|l|}
    \hline
    space ID & kernel height & kernel width & number of kernel              & stride height & stride width & pool size \\ \hline
    full space        & 1, 3, 5, 7, 9  & 1, 3, 5, 7, 9 & 4, 8, 12, 24, 36, 64, 128 & 1, 2, 3, 4, 5  & 1, 2, 3, 4, 5 & 1, 2       \\ \hline
    subspace S1        & 1, 5, 7        & 3, 5, 7       & 24, 36, 48, 64            & 1, 2, 3        & 1, 2, 3       & 1, 2       \\ \hline
    subspace S2        & 1, 3, 5, 7     & 1, 3, 5, 7    & 24, 36, 48, 64            & 1, 2, 3        & 1, 2, 3       & 1, 2       \\ \hline
    subspace S3        & 1, 3, 5, 7, 9  & 1, 3, 5, 7, 9 & 4, 8, 12, 24, 36, 64, 128 & 0, 1, 2, 3     & 0, 1, 2, 3    & 1          \\ \hline
    subspace S4        & 1, 3, 5, 7, 9  & 1, 3, 5, 7, 9 & 4, 8, 12, 24, 36, 64, 128 & 1, 2, 3, 4, 5  & 1, 2, 3, 4, 5 & 1          \\ \hline
    subspace S5        & 1, 3, 5, 7, 9  & 1, 3, 5, 7, 9 & 4, 8, 12, 24, 36, 64, 128 & 1, 2, 3, 4, 5  & 1, 2, 3, 4, 5 & 1          \\ \hline
    subspace S6        & 1, 3, 5        & 1, 3, 5       & 4, 8, 12                  & 1, 2, 3        & 1, 2, 3       & 1          \\ \hline
    subspace S7        & 5, 7, 9        & 5, 7, 9       & 32, 64, 128               & 3, 4, 5        & 3, 4, 5       & 1          \\ \hline
    subspace S8        & 5, 7, 9        & 5, 7, 9       & 32, 64, 128               & 3, 4, 5        & 3, 4, 5       & 1          \\ \hline
    subspace S9        & 1, 3, 5        & 1, 3, 5       & 24, 36                    & 1, 2, 3        & 1, 2, 3       & 1          \\ \hline

    \end{tabular}
    }
    \label{tab:arch}
\end{table*}

\begin{table*}[ht]
\centering
\caption{Model quantization subspaces for subspaces S1 to S9 and full space.}
\resizebox{\columnwidth}{!}{%
    \begin{tabular}{|l|l|l|l|l|}
    \hline
    space ID & act\_num\_int\_bits & act\_num\_frac\_bits & weight\_num\_int\_bits & weight\_num\_frac\_bits \\ \hline
    full space        & 0, 1, 2, 3          & 0, 1, 2, 3, 4, 5, 6  & 0, 1, 2, 3, 4          & 0, 1, 2, 3, 4, 5, 6     \\ \hline
    subspace S1        & 1, 2, 3             & 1, 2, 3, 4, 5        & 0, 1, 2, 3, 4          & 2, 3, 4, 5              \\ \hline
    subspace S2        & 0, 1, 2, 3          & 0, 1, 2, 3, 4, 5, 6  & 0, 1, 2, 3             & 0, 1, 2, 3, 4, 5, 6     \\ \hline
    subspace S3        & 0, 1, 2, 3          & 0, 1, 2, 3, 4, 5, 6  & 0, 1, 2, 3             & 0, 1, 2, 3, 4, 5, 6     \\ \hline
    subspace S4        & 2, 3                & 4, 5, 6              & 2, 3                   & 4, 5, 6                 \\ \hline
    subspace S5        & 0, 1                & 1, 2, 3              & 0, 1                   & 1, 2, 3                 \\ \hline
    subspace S6        & 0, 1, 2, 3          & 0, 1, 2, 3, 4, 5, 6  & 0, 1, 2, 3             & 0, 1, 2, 3, 4, 5, 6     \\ \hline
    subspace S7        & 0, 1, 2, 3          & 0, 1, 2, 3, 4, 5, 6  & 0, 1, 2, 3             & 0, 1, 2, 3, 4, 5, 6     \\ \hline
    subspace S8        & 2, 3                & 4, 5, 6              & 2, 3                   & 4, 5, 6                 \\ \hline
    subspace S9        & 2, 3                & 5, 6                 & 2, 3                   & 5, 6                    \\ \hline

    \end{tabular}
    }
    \label{tab:quan}
\end{table*}

In this section, we will firstly introduce the experimental setup in subsection \ref{sec:sub:experiment_setup}, including the NAS setting general information, the hardware information, and the weak miner simulation. 
In subsection \ref{sec:sub:benchmark}, we will analyze the performance of the mining pool searching in nine different subspaces versus the performance of an individual miner searching in the full space.
In subsection \ref{sec:sub:miners_reliability_evaluation}, we will analyze the effect reliability of miners on the performance of the mining pool. 

\subsection{Experiment setup} \label{sec:sub:experiment_setup}
\subsubsection{Experiment overview}
To evaluate the performance of the mining pool, we adopt the popular RNN based NAS \cite{zoph2016neural_replace}. The NAS algorithm is targeting to find a CNN architecture for classification tasks under a certain hardware constrain. We use CIFAR-10 dataset \cite{krizhevsky2009cifar} which has 10 classes images. All experiments were deployed on the workstation with with Intel(R) Core(TM) i7-9900K CPU @ 3.60GHz, 32Gb RAM, 2$\times$ GTX 1080 Ti. 

\subsubsection{Hardware constraints}
The hardware constraints include the size of the lookup table of FPGA, and throughput. The size of the lookup table of FPGA represents the required chip size. In this NAS task, we will give 100,000 as of the upper bound. Throughput also represents the latency of the architecture. Here, we set 10 as the lower bound. Due to these strong hardware constraints, the performance of neural architecture may not be able to beat the performance of a model with unlimited hardware resources. 

\subsubsection{Searching and reward}
In the searching process, we will first evaluate the required chip size and latency. Based on the hyperparameter, if the neural architecture is not exceeding the bounds, the miner will start training and return the best testing accuracy within 30 epochs of training in this experiment. If the result beats the previous best reward, it will be updated as the best reward. Here, the reward means the testing accuracy for the image classifier. The same value is named reward for the RNN. This RNN is for selecting the best hyperparameter value in the given searching range.
If the required lookup table or throughput exceeding the bound, the controller will return zero as the reward for RNN. For each subspace, the controller will search for 2000 episodes and each episode will train for 30 epochs. 

\subsubsection{Experiments input data}
In the table \ref{tab:arch}, it shows the model architecture subspaces for subspaces S1 to S9 and full space. In table \ref{tab:quan}, it shows the model quantization subspaces for subspaces S1 to S9 and full space. The hyperparameters for model architecture include kernel height, kernel width, number of kernels, stride height, stride width, and pool size. The hyperparameters for model quantization include the number of bits of the activation integer part, number of bits of activation fraction part, number of bits of the weight integer part, and number of bits of weight fraction part. 

The total number of hyperparameter $'n'$ equals to 10. The maximum sizes $'k'$ of the searching ranges $'r_i'$ are (5, 5, 7, 5, 5, 2, 4, 7, 5, 7). The total number of miners $'m'$ equals to 9. The full searching ranges for each hyperparameter are given in the row of full space in table \ref{tab:arch} and \ref{tab:quan}.

For the rows of subspaces S1 to S9, it gives assigned searching ranges of each hyperparameter of each subspace. In the algorithm \ref{algo:subspace}, The table $T$ saves all subset of each searching range. Here, $T[i]$ saves a list of all subset of the searching range of the $i-th$ hyperparameter. Because the table $T$ contains all subset and miners will fetch searching range for a certain hyperparameter randomly, it is possible to achieve one subspace overlap another subspace. This algorithm \ref{algo:subspace} may lead to inefficient searching space partition. But we will demonstrate that the searching space overlapping is not an efficiency issue in subsection \ref{sec:sub:benchmark}. An efficient searching space partition may improve NAS mining pool performance, but it is out of our current scope.  

\subsubsection{Weak miner simulation} 
As explained in section \ref{sec:methodology}, the weak miner may drawback the overall performance, thus we will need to simulated the weak miners in our experiment. In practice, many reasons will affect the performance of a miner, such as frequency and cores number of CPU, size, and bandwidth of memory, GPU, etc.. This will bring difficulties for us to evaluate the consequence when we assign the same amount of workload. We will use the workstation with the same configuration and hardware, but only allow a shorter run time to simulate a weak miner, such machine will run 1/10 of the total strong miner run time to simulate 10 times weaker miner.

\subsection{benchmark of mining pool} \label{sec:sub:benchmark}
\begin{figure*}[ht]
\begin{center}
 \includegraphics[width=1\columnwidth]{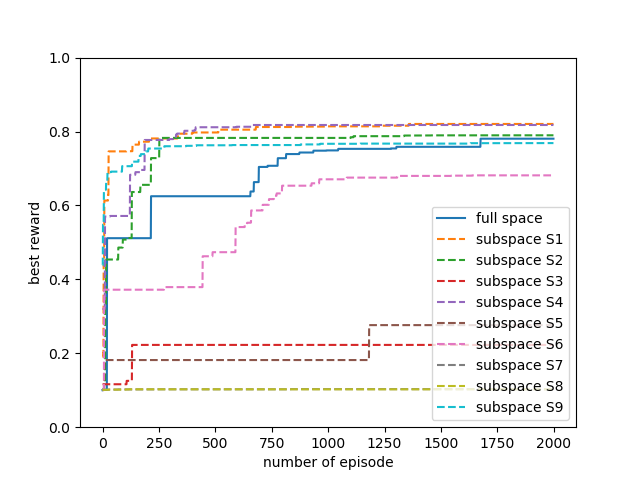}
\end{center} 
  \caption{The results of NAS in full searching space, and subspace 1 to 9. The x-axis is the number of episodes and y-axis is the best rewards.}
\label{fig:full}
\end{figure*}

The experiments conduct as it is introduced in subsection \ref{sec:sub:experiment_setup}. As an individual miner search neural architecture in a searching space, it evaluates the required sizes of the lookup table of FPGA. If the required size is bigger than the given upper bound, the miner returns zero as the accuracy of the current configuration. Otherwise, the miner continues to evaluate the throughput of the current configuration. If it is lower than the given lower bound, it will return zero as the accuracy of the current configuration. Otherwise, the miner will start to train the current neural architecture for 30 epochs. Then the controller will test the model. If the result beats the current best reward, the controller updates the value of the current best reward. Otherwise, the controller only saves the testing accuracy as the reward of the current configuration. The controller repeats this procedure for 2000 episodes. 

 The current reward is the test accuracy of the current configuration and it is for training the RNN model. After the performance of the RNN model improved, the controller will find better performance neural architecture under the given hardware constraints more efficiently. 

In the novel consensuses blockchain system, the full nodes only check the best performance models. Therefore, we will only shows the best reward (accuracy) in the Fig. \ref{fig:full} and \ref{fig:1249}. 
In the Fig. \ref{fig:full}, the solid line shows the result of one individual miner searching in the full searching space. The dash lines show the results of nine miners searching in the searching subspaces S1 to S9 independently. In the Fig. \ref{fig:full}, four of the miners return low-performance results, and four of the miners return high-performance results. Here, the results for subspace S7 and subspace S8 are overlapping. 

The searching ranges of each hyperparameter is given in the row of full space in the table \ref{tab:arch} and table \ref{tab:quan}. 
In this subsection, we will evaluate whether this mining pool will help to find a better architecture than a single miner. This seems to be straightforward. Due to the network latency and system overhead, it is not always true that a group of machines find a better neural network architecture than a single machine. In our design, we applied embarrassingly parallel computation parallelism and there is no dependence between different miners. Therefore, the network latency will not strongly affect performance.

As shown in the Fig. \ref{fig:full}, this embarrassingly parallel computation parallelism helps the whole mining pool to be more competitive than a single miner.
Four miners are searching in subspace S1, S2, S4, and S9 from the mining pool to return better results than the individual miner during the majority of the mining time. Three miners return better results at the end, and two miners return better results from beginning to the end. The miner searching in subspaces S1 finds very good performance neural architecture at a very early time and finds the best performance neural architecture at the end. In the Fig. \ref{fig:1249}, it is a zoom-in result where the y-axis range is from 0.7 to 0.9. Around episode 200 to 250, the miner searching subspace S1 is overtaken by the miners in subspace S2 and S4. The miner searching subspace S4 keeps the leading position till episode 1300 to 1400. In practice, miners will adjust the number of episodes due to the length of the given training phase and their computation power. These results show that more miners searching in different subspaces will help the mining pool to be more competitive during the whole training phase.

\subsection{Evaluation for effect of miners reliability} \label{sec:sub:miners_reliability_evaluation}

\subsubsection{Miners reliability}
\begin{figure*}[ht]
\begin{center}
 \includegraphics[width=1\columnwidth]{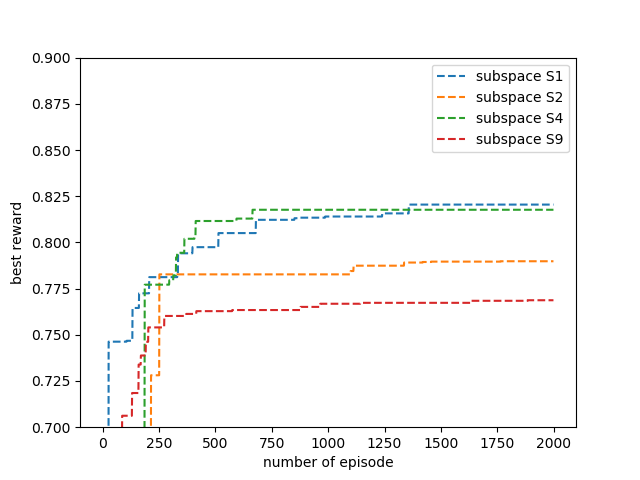}
\end{center} 
  \caption{The results of NAS in subspace S1, S2, S4, and S9. The x-axis is the number of episodes and y-axis is the best rewards with the range from 0.7 to 0.9.}
\label{fig:1249}
\end{figure*}

In subsection \ref{sec:sub:benchmark}, it shows the performance of this mining pool based on embarrassingly parallel computations is more competitive than the performance of an individual miner. In this small scale of evaluation, only 9 miners are searching in 9 different subspaces. In the Fig. \ref{fig:full}, four of the miners return low-performance results, and four of the miners return high-performance results. The high performance versus low-performance ratio is not promised in a large scale of testing, but it shows a good amount of miners return high-performance results. Especially, the results from different subspace S1 and S4 are very close in the Fig. \ref{fig:1249}. In the large scale mining pool, if there are multiple miners like the miners in subspace S1 and S4. It means the overall performance of the mining pool is not depends on any individual miner. 

\begin{figure*}[ht]
\begin{center}
 \includegraphics[width=1\columnwidth]{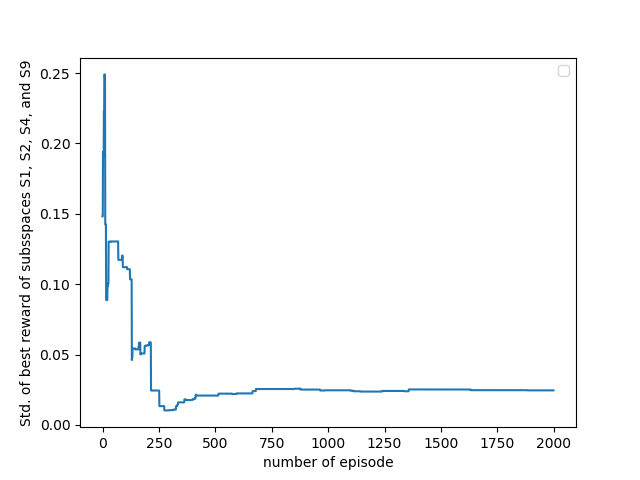}
\end{center} 
  \caption{The standard deviation of the results for the miners searching in subspace S1, S2, S4, and S9. The x-axis is the number of episodes and y-axis is the standard deviation value.}
\label{fig:std1249}
\end{figure*}

In the Fig. \ref{fig:std1249}, it shows the standard deviation of the results for the miners searching in subspaces S1, S2, S4, and S9. This standard deviation is calculated from all high reward results from the current experiments. The standard deviation is high when the number of the episode is small. After 500 episodes, the standard deviation value will become low and flatten after more episodes of searching. When we need to evaluate the mining pool performance it is also important to check whether the overall performance of the mining pool depends on any individual miners. The data should only be collected based on all the miners who returns high reward values. Once the standard deviation value amount these miners raise, it is necessary for the mining pool manager to prepare backup miners to continue the searching task if any high reward miners leave the pool. The backup miners can be selected from the miners who returns low reward. In the Fig. \ref{fig:full}, the miners searching subspaces S1, S2, S4, and S9 are considered as high reward miners, and the miners searching subspaces S3, S5, S7, and S8 are considered as low reward miners in this experiment. Only strong miners search for neural architecture. Here, both high and low reward miners are strong miners. Because miners may leave the mining pool at any time, the manager needs to handle this case.

\subsubsection{Exploration and exploiting} 
As it is described in section \ref{sec:methodology}, the mining pool manager assigns exploration tasks to strong miners and assign exploiting tasks to weak miners once strong miners return confirmed neural architecture. Therefore, there is no chance that a weak miner will drawback the overall performance of the mining pool and the strong miners will be able to spend the valuable computation power on searching in more space. Thus, the performance could be maximum. 

As it is explained in section \ref{sec:methodology}, a weak miner may drawback the overall performance. As description in subsection \ref{sec:sub:experiment_setup}, we will simulate a weak miner by scale the run time down to 1/10 of the strong miner. When the weak miner is assigned to search architecture from space two, it is the most possible setting to find the best solution. 
Based on the Fig. \ref{fig:full} and \ref{fig:1249}, it easy to find high performance neural architecture in the subspaces S1 and S4. 
Without this exploration and exploiting strategy, if 10 times slower miners are assigned to search subspaces S1 and S4, the miners in subspace s2 will be the only contributor to the overall best performance of the mining pool which is 0.7829. The weak miners will return 0.5701 and 0.7724. With exploration and exploiting strategy, the overall performance of the mining pool is 0.8204. 
Therefore this weak miner simulation shows that the weak miner only able to search for 200 episodes and weak miners will eventually waste a good opportunity to find the best solution. Although other miners may achieve similar results to the best solution from different subspaces if the mining pool manager will not schedule other miners to search in the same subspace. Therefore, the weak miner may waste a good opportunity and drawback the performance of the whole mining pool without this exploration and exploiting strategy.


\section{Conclusion} \label{sec:conclusion}


In this paper, we briefly introduced recent PoUW novel consensuses and details of the consensuses based on deep learning training. We adopt NAS as the workload to demonstrate the concepts. For a public accessible cryptocurrency blockchain system, earning tokens is the incentive for an individual miner to participates in mining. In a large-scale cryptocurrency system, the possibility of earning tokens for an individual is very low. A mining pool will support individual miners to be profitable in the large-scale cryptocurrency blockchain system.
In this project, we are the first to demonstrate a mining pool solution to support novel deep learning-based consensuses. 

In section \ref{sec:methodology}, we have introduced the function of pool manager, subspaces partition algorithm, and exploration \& exploitation strategy. In section \ref{sec:experiment}, we evaluate the performance of the mining pool, and it is more competitive than an individual miner who searches the full space individually. Due to the uncertainty of individual miners, the mining pool manager will keep track of the high-performance miners and prepare backup miners as introduced in subsection \ref{sec:sub:miners_reliability_evaluation}.

